# STREAM OF DARK MATTER AS A POSSIBLE CAUSE OF THE OPERA CLOCKS' SYNCHRONIZATION SIGNALS DELAY

Jean Paul Mbelek[1]

1. Sangha Center for Astronomy, Astrophysics and Cosmology, Sangha, Mali

Correspondence to: Jean Paul Mbelek[1] Correspondence and requests for materials should be addressed to J.P.M. (Email: mbelek.jp@wanadoo.fr).

Abstract : A stream of weakly interacting massive particles (WIMPs) gravitationally scattered outwards within the Earth yields a delay, $\delta t \approx 60$ ns, in good agreement with the results of the OPERA experiment. Conversely, the OPERA experiment may be seen as the unveiling of the first hint of a dark matter particle probed with the photons of the GPS communication signals and using the neutrino beam of the CNGS as a velocity standard. Our analysis yields the true neutrino velocity, $V_\nu$, less than the speed of light in vacuum, one finds $(V_\nu - c)/c = -(1.8 \pm 0.4) \times 10^{-9}$. A new experimental test still predicting $\delta t \sim 60$ ns instead of $\delta t \sim 0.6$ µs is suggested, based on the use of the long baseline of the order 7 800 km between either the Fermilab neutrino production site and the OPERA detector or the CERN neutrino production site and the MINOS detector.

I. Introduction

The OPERA collaboration has recently ruled out the uncertainties due to the duration of the CERN SPS (Super Proton Synchrotron) proton pulses and thereby confirming the claim [1] that their precise measurement of the time of flight (TOF) of neutrinos and the distance, $d = (730534.61 \pm 0.20)$ m, from the CERN to the Gran Sasso imply for the apparent velocity, $v$, of muon neutrinos the estimate $\delta = (v - c)/c = (2.37 \pm 0.32 \text{ (stat.)} ^{+0.34}_{-0.24} \text{ (sys.)}) \times 10^{-5}$. The above result is derived subsequent to the measurement at 6σ level of a putative early arrival time, $\delta t = (57.8 \pm 7.8 \text{ (stat.)}^{+8.3}_{-5.9} \text{ (sys.)})$ ns, of muon neutrinos with respect to the one computed by relying on the GPS communication signals for clocks synchronization. However, we argue that the faster than light interpretation of the OPERA experiment is definitely not satisfactory, if the conservation of energy-momentum still holds. Indeed, in order to determine an accurate position from range data, one needs to account for all the propagation effects and time offset, in addition to change of refractive delay along path length and change of path length [2]. Now, one may argue that the contribution of charged carriers (*e.g.*, the solar wind) to the average refractive index of the propagation medium of the GPS communication signals, $n_0$, have already been taken into account through the relation $n_0 = (1 + \chi_e)^{1/2}$ involved by the induced dielectric polarization density $\mathbf{P} = \varepsilon_0 \chi_e \mathbf{E}$, so that the value of the average refractive index used in the simulations identifies to $n_0$. However, the latter argument should not hold anymore by considering alternatively a stream of WIMPs ("wimp wind") [3-6], since these dark matter particles are not composite and electrically neutral. Indeed, such uncharged particles are not subject to the geomagnetic field and bring no contribution either to the electric susceptibility, $\chi_e$, or the magnetic susceptibility, $\chi_m$. Besides, we show that the faster than light interpretation of the OPERA experiment is at odds with the conservation laws for energy and momentum. Instead, our interpretation of the time delay, $\delta t$, of the OPERA experiment is based on the velocity addition law of special relativity (SR). In this respect, this is achieved by correcting for the first order term in V/c that accounts for the flow velocity, V, of a stream of galactic WIMPs gravitationally scattered outwards within the Earth. As a consequence, the OPERA experiment turns out to be consistent with SR in as much as it is actually the GPS communication signals that propagate more slowly than originally expected from simulations.

II. Internal inconsistency of the superluminal neutrinos interpretation of the OPERA result

Let us consider the $\nu_\mu$ and $\mu^+$ leptons obtained from the decay of either the $\pi^+$ or $K^+$ mesons produced by the collision of the CERN SPS protons with the graphite target. The conservation laws for energy and momentum read

$P_p = P_\nu + P_X + P_Y$ (1)

$E_p + m_Y c^2 = E_\nu + E_X + E_Y$ (2)

$E_p^2 = P_p^2 c^2 + m_p^2 c^4$ (3)

$E_X^2 = P_X^2 c^2 + m_X^2 c^4$ (4)

$E_Y^2 = P_Y^2 c^2 + m_Y^2 c^4$ (5)

$P_p = E_p V_p/c^2$, $P_X = E_X V_X/c^2$ and $P_Y = E_Y V_Y/c^2$ (6)

where the subscript X stands for the $\mu^+$ meson and Y stands for a carbon atom or ion of the graphite target. Since the velocity of the CNGS neutrinos only differs slightly from the velocity of light in vacuum, the SR relation still holds at least as a first approximation, namely

$P_\nu = E_\nu V_\nu/c^2$ (7)

Combining relations (1-7) by taking into account the conditions $E_\nu < P_p c < E_p$ and assuming a superluminal neutrino, namely $V_\nu > c$, would yield

$(E_p - P_p c) E_\nu + E_X E_Y [1 - (V_X V_Y/c^2)] > \frac{1}{2} (m_p^2 + m_\nu^2 - m_X^2) c^4$. (8)

Since $E_X \sim E_\nu$, $E_Y \ll E_p - P_p c$, $V_X \leq c$, $V_Y \ll c$, the above inequality implies

$E_\nu > (m_p^2 + m_\nu^2 - m_X^2) c^4/[2(E_p - P_p c)] > [1 - (m_X/m_p)^2] \times P_p c$. (9)

Now, the CNGS beam is produced by protons with a momentum $P_p$ = 400 GeV/c, and the masses of the proton and the pion are respectively $m_p$ = 0,94 GeV/$c^2$ and $m_\mu$ = 0,11 GeV/$c^2$. Hence, according to the above inequality the energy, $E_\nu$, of the CNGS neutrinos should be greater than 395 GeV. Now, the energy range of the CNGS neutrinos is only 13.9 GeV - 42.9 GeV with the average energy equal to 17 GeV which is about twenty times smaller than the values requirred by the above inequality. So, the latter analysis shows clearly that the faster than light interpretation of the OPERA result leads to an internal inconsistency and as such should be rejected. Moreover, on the experimental ground by referring to the Cohen and Glashow prediction [7] (see also [8]), the ICARUS team has refuted the superluminal interpretation of the OPERA result [9] and a recent reinterpretation of the NOMAD results points to the same conclusion too [10].

III. GPS communication signals propagating more slowly than expected from simulations

Let $V_\nu$ and $TOF_\nu = d/V_\nu$ be respectively the true velocity and the TOF of the muon neutrinos in the Earth's crust with respect to the OPERA laboratories, where d denotes the distance between the target focal point and the origin of the OPERA reference frame. Now, the TOF of the photons travelling the same distance, d, in the vacuum would be $TOF_c = d/c$. However, the TOF of the photons, $TOF'_c$, is actually determined by simulations. Whence, the equality

$$V_\nu \, TOF_\nu = c \, TOF'_c. \quad (10)$$

We emphasize that the nominal value $TOF'_c$ is derived by using a fiducial mean refractive index, $n_0$, and a fiducial velocity, $V_{\gamma/0} = c/n_0$, for the GPS communication signals crossing the magnetosphere down to the atmosphere. Now, in the terrestrial reference frame, the SR velocity addition law implies that light with initial speed $V_{\gamma/0}$ slows down to the speed $V_\gamma = ||(\mathbf{V}_{\gamma/0} + \mathbf{V}) \times [1 + (\mathbf{V}_{\gamma/0} \cdot \mathbf{V}/c^2)]^{-1}|| \approx (V_{\gamma/0} - V) \times [1 - (V_{\gamma/0} V/c^2)]^{-1}$ by crossing a dispersive medium moving with velocity $\mathbf{V}$ opposite to $\mathbf{V}_{\gamma/0}$. The same effect applies to the neutrinos in the crust of the Earth but without any noticeable change of speed, since $d \ll D$, $V \ll c$ and their track is almost perpendicular to the Earth's radius and hence to $\mathbf{V}$. Thus, the stream of WIMPs scattered outwards within the Earth will perturb $V_\gamma$ so that the effective refractive index increases slightly to $n = c/V_\gamma$ and the TOF of the photons lengthens to $TOF_c > TOF'_c$. It results in a delay in excess proportional to the distance, for the photons of the GPS communication signals. Let $\delta t$ be the delay in excess as compared to what would be expected while the GPS communication signals travel a distance D from the magnetosphere down to the surface of the Earth. This translates into a delay, $\delta t_{\gamma/\nu}$, in the TOF of the photons as compared to that of the neutrinos travelling the distance d within the Earth's crust. Thus, the counterpart to relation (10) for the photons of the GPS communication signals reads

$$V_\gamma \, [TOF_\nu + \delta t_{\gamma/\nu}] = V_{\gamma/0} \, TOF'_c \quad (11)$$

or otherwise stated

$$v \, TOF_\nu = c \, TOF_c, \quad (12)$$

where $v = c + (\delta t/\delta t_{\gamma/\nu}) [(n/n_0) V_\nu - c]$ is interpreted as the apparent velocity of the neutrinos with respect to the OPERA experiment laboratories. Since $D > d$, $n > n_0$ and $V_\nu \approx c$ with great precision, it follows $v > c$ which may lead to conclude wrongly to superluminal neutrinos such that

$$(v - c)/c = \delta t/TOF_\nu \approx \delta t/(TOF'_c - \delta t), \quad (13)$$

and $\delta t = TOF_c - TOF_\nu$ is improperly interpreted as an early arrival time of the muon neutrinos as compared with the TOF computed assuming the speed of light in vacuum.

Thus, by taking into account the correction to the TOF of the photons due to the unaccounted flow of WIMPs, the one-way delay in excess of the GPS communication signals reads to the first-order approximation in V/c,

$$\delta t = (D/V_\gamma) - (D/V_{\gamma/0}) = (n - n_0) \times (D/c) \approx 2(n_0 - 1) DV/c^2. \qquad (14)$$

where

$$n = [n_0 - (V/c)] \times [1 - (n_0 V/c)]^{-1} \approx n_0 + 2(n_0 - 1) \times (V/c). \qquad (15)$$

Clearly, relation (14) is energy independent. By using the value $n_0 - 1 = 2.926 \times 10^{-4}$ found in text books, $D = (x^2 + z^2)^{1/2} \approx z = 20\,200$ km ($0 \leq x \leq d$ ; $z$ = altitude of the GPS satellites) and taking the average speed of the WIMPs scattered by the Earth equal to the typical value $V \approx (440 \pm 105)$ km/s, one finds a time delay in excess $\delta t$ in good agreement with the OPERA result. Since $\delta t_{\gamma/\nu}$ and $\delta t$ are both proportional to their corresponding distance, it follows $\delta t_{\gamma/\nu} = (d/D)\,\delta t$, so that combining relations (14) and (15) yields

$$v = c + (D/d)\,[(n/n_0)\,V_\nu - c] \qquad (16)$$

and consequently,

$$(v - c)/c = (D/d)\,[(n/n_0) \times (V_\nu/c) - 1] \approx (D/d)\,(n - n_0) \approx c\,\delta t/d. \qquad (17)$$

Relations (14-17) above provide the true neutrino velocity, $V_\nu$. One finds

$$(V_\nu - c)/c \approx -2(n_0 - 1) \times (V/c) + \delta\,(d/D) = -(1.8 \pm 0.4) \times 10^{-9}, \qquad (18)$$

which is just less than the speed of light in vacuum, c, in conformity with Einstein causality and neutrino oscillations that imply nonzero neutrino masses.

IV. Conclusion

We have investigated the claim for superluminal neutrinos dealing with the OPERA experiment. In view of our analysis based on the energy-momentum conservation, we are led to conclude that the effect unveiled by the OPERA experiment does not really conflict with Einstein causality. Instead of faster than light neutrinos, it fits quite well with light, namely the GPS communication signals, slower than would be expected. Hence, we call for experiments between either the CERN SPS and the MINOS detector or between the Fermilab and the OPERA detector so that the baseline of the neutrino beam becomes ten times greater (about 7800 km) than that of the present OPERA experiment. So, if our prediction holds, one should find for the latter experiments an energy independent time delay $\delta t \sim (60 \pm 10)$ ns instead of $\delta t \sim (0.6 \pm 0.1)$ μs.